\documentclass[econometrics,article,submit,moreauthors,pdftex]{mdpi} 

\firstpage{1} 
\pubvolume{xx}
\issuenum{1}
\articlenumber{5}
\pubyear{2019}
\copyrightyear{2019}
\history{Received: date; Accepted: date; Published: date}
\usepackage{bm}
\usepackage{multirow}

\Title{Bayesian model averaging with the integrated nested Laplace approximation}

\Author{Virgilio G\'omez-Rubio, $^{1,\dagger,\ddagger}$*\orcidA{}, Roger S. Bivand $^{2,\ddagger}$\orcidB{} and H\aa{}vard Rue$^{3,\ddagger}$}

\AuthorNames{Virgilio G\'omez-Rubio, Roger S. Bivand and H\aa{}vard Rue}

\address{%
$^{1}$ \quad Department of Mathematics, Universidad de Castilla-La Mancha, Albacete, Spain; Virgilio.Gomez@uclm.es\\
$^{2}$ \quad Department of Economics, Norwegian School of Economics, Bergen, Norway; roger.bivand@nhh.no\\
$^{3}$ \quad CEMSE Division, King Abdullah University of Science and Technology, Thuwal, Saudi Arabia; haavard.rue@kaust.edu.sa}

\corres{Correspondence: Virgilio.Gomez@uclm.es; Tel.: +34-967-59-92-00, ext. 8291}

\firstnote{Current address: Department of Mathematics, School of Industrial Engineering, Universidad de Castilla-La mancha, Avda. Espa\~na s/n, 02071 Albacete (Spain)}
\secondnote{These authors contributed equally to this work.}
\abstract{The integrated nested Laplace approximation (INLA) for Bayesian 
inference is an efficient approach to estimate the posterior marginal
distributions of the parameters and latent effects of Bayesian hierarchical
models that can be expressed as latent Gaussian Markov random fields (GMRF).
The representation as a GMRF allows the associated software R-INLA to estimate
the posterior marginals in a fraction of the time as typical Markov chain Monte
Carlo algorithms. INLA can be extended by means of Bayesian model averaging
(BMA) to increase the number of models that it can fit to conditional latent
GMRF. In
this paper we review the use of BMA with INLA and propose a new example on
spatial econometrics models.
}


\begin{document}
%%%%%%%%%%%%%%%%%%%%%%%%%%%%%%%%%%%%%%%%%%

%%%%%%%%%%%%%%%%%%%%%%%%%%%%%%%%%%%%%%%%%%

\section{Introduction}

Bayesian model averaging \citep[BMA, see, for example,][]{Hoetingetal:1999} is
a way to combine different Bayesian hierarchical models that can be used to
estimate highly parameterized models.  By computing an average model, the
uncertainty about the model choice is taken into account when estimating the
uncertainty of the model parameters.

As BMA often requires fitting a large number of models, this can be time
consuming when the time required to fit each of the models is large.  The
integrated nested Laplace approximation \citep[INLA,][]{Rueetal:2009} offers an
alternative to computationally intensive Markov chain Monte Carlo
\citep[MCMC,][]{Gilksetal:1996} methods. INLA focuses on obtaining an
approximation of the posterior marginal distributions of the models parameters
of latent GMRF models \citep{RueHeld:2005}. Hence, BMA with INLA is based on
combining the resulting marginals from all the models averaged.

\cite{Bivandetal:2014,Bivandetal:2015} use this approach to fit some spatial
econometrics models by fitting conditional models on some of the
hyperparameters of the original models. The resulting models are then combined
using BMA to obtain estimates of the marginals of the hyperparameters of the
original model.  \cite{GomezRubioRue:2018} have embedded INLA within MCMC so
that the joint posterior distribution of a subset of the model parameters is
estimated using the Metropolis-Hastings algorithm \citep{Metropolisetal:1953,Hastings:1970}. This requires
fitting a model with INLA (conditional on some parameters) at each step, so
that the resulting models can also be combined to obtain the posterior
marginals of the remainder of the model parameters.

BMA with INLA relies on a weighted sum of the conditional marginals obtained
from a family of conditional models on some hyperparameters.  Weights are
computed by using Bayes' rule and they depend on the marginal likelihood of the
conditional model and the prior distribution of the conditioning
hyperparameters.  This new approach is described in detail in
Section~\ref{sec:BMA}. We will illustrate how this works by developing an
example on spatial econometrics models described in Section~\ref{sec:SE}.

This paper is organized as follows.  Spatial econometrics models are summarized
in Section~\ref{sec:SE}.  Next, an introduction to INLA is given in
Section~\ref{sec:INLA}.  This is followed by a description of Bayesian model
averaging (with INLA) in Section~\ref{sec:BMA}.  An example is developed in
Section~\ref{sec:example}. Finally, Section~\ref{sec:discussion} gives a
summary of the paper and includes a discussion on the main results.

\section{Spatial econometrics models}
\label{sec:SE}

Spatial econometrics models \citep{LeSagePace:2009} are often employed to
account for spatial autocorrelation in the data. Usually, these models include
one or more spatial autoregressive terms. \citet{Manski:1993} propose a model
that includes an autoregressive term on the response and another one in the
error term:

$$
y = \rho W y + X \beta + W X \gamma + u
$$

Here, $y$ is the response, $X$ a matrix of covariates with coefficients
$\beta$, $W$ an adjacency matrix and $W X$ are lagged covariates with
coefficients $\gamma$.  Finally, $u$ is an error term. This error term is
often modeled to include spatial autocorrelation:

$$
u = \lambda W u + e
$$

Here, $e$ is a vector of Gaussian observations with zero mean
and precision $\tau$.

The previous model can be rewritten as

$$
y = (I - \rho W)^{-1} (\beta X) + u^{'}
$$

with $u^{'}$ an error term with a Gaussian distribution with
zero mean and precision matrix 

$$
\tau (I - \rho W^{\top}) (I - \lambda W^{\top}) (I - \lambda W) (I - \rho W)
$$

Note that here the same adjacency matrix $W$ has been used for the two
autocorrelated terms, but different adjacency matrices could be used.
The range of $\rho$ and $\lambda$ is determined by the eigenvalues 
of $W$. When $W$ is taken to be row-standardized, the range is
the interval $(1/m, 1))$, where $m$ is the minimum eigenvalue of $W$ \citep[see, for example,][]{Haining:2003}. In this case, the lagged covariates
$W X$ represent the average value at the neighbors, which is useful
when interpreting the results.

In a Bayesian context, a prior needs to be set on every model parameter.  For
the spatial autocorrelation parameter a uniform in the interval $(-1, 1)$ will
be used. For the coefficients of the covariates, a Normal with zero mean and
precision 1000 is used, and for precision $\tau$ Gamma with parameters $0.01$
and $0.01$.

This model is often referred to as spatial autoregressive combined (SAC) model.
Other important models in spatial econometrics appear when some of the terms in
the SAC model are dropped. See, for example, \cite{GomezRubioetal:2017} and how
these models are fit with INLA.

In spatial econometrics models it is of interest how changes in the value of
the covariates affect the response in neighboring regions. These spill-over
effects or \textit{impacts} \citep{LeSagePace:2009} are caused by the term $(I - \rho W )^{-1}$ that
multiplies the covariates and they are defined as

$$
\frac{\partial y_{i}}{\partial x_{jk}},\ i,j=1,\ldots,n; \ k = 1, \ldots, p
$$
\noindent
where $n$ is the number of observations and $p$ the number of covariates 
and $x_{jk}$ is the value of covariate $k$ in region $j$.

Hence, for each covariate $k$ there will be an associated matrix of impacts.
The diagonal values are known as \textit{direct} impacts as they
measure the effect of changing a covariate on the same areas.  The
off-diagonal values are known as \textit{indirect} impacts as they
measure the change of the response at neighboring areas when covariate $k$
changes. Finally, \textit{total} impacts are the sum of direct and indirect
impacts.

\cite{GomezRubioetal:2017} describe how the impacts are for different
spatial econometrics models. For the SAC model, the impacts matrix
for covariate $k$ are 

$$
S_r(W) = (I - \rho W)^{-1} I \beta_r
$$

In practice, average impacts are reported as a summary of the direct, indirect
and total impacts \citep[for details see, for example, ][]{LeSagePace:2009,GomezRubioetal:2017}. In particular, the average direct impact is the trace
of $S_r(W)$ divided by $n$, the average indirect impact is defined as 
the sum of the off-diagonal elements divided by $n$ and 
the average total impact is the sum of all elements of $S_r(W)$ divided by $n$.
Average total impact is also the sum of the average direct and average indirect
impacts.

For the SAC model, the average total impact is $\beta_r / (1 - \rho)$ and the
average direct impact is $n^{-1} tr((I - \rho W)^{-1})\beta_r$. The average
indirect impact can be computed as the difference between the average total
and average direct impacts, $[1 / (1 - \rho) - n^{-1} tr((I - \rho
W)^{-1})]\beta_r$,  or by computing the sum of the off-diagonal elements
of $S_r(W)$ divided by $n$.

Note that computing the impacts depends on parameters $\rho$ and $\beta_r$. For
this reason, the joint posterior distribution of both parameters would be
required.  A explained below, this is not a problem because it can be rewritten
as $\pi(\beta_r, \rho \mid \mathbf{y}) = \pi(\beta_r \mid \mathbf{y}, \rho)
\pi(\rho \mid \mathbf{y})$.

\section{The integrated nested Laplace approximation}
\label{sec:INLA}

Markov chain Monte Carlo algorithms \citep[see, for example,][]{Brooksetal:2011}
are typically used to estimate the joint posterior distribution of the
ensemble of parameters and latent effects of a hierarchical Bayesian model.
However, these algorithms are often slow when dealing with high parameterized
models.

Recently \citet{Rueetal:2009}, have proposed a computationally efficient
numerical to estimate the posterior marginals of the hyperparameters and latent
effects. This method is called integrated nested Laplace approximation (INLA)
because it is based on repeatedly using the Laplace approximation to estimate
the posterior marginals.  In addition, INLA focuses on models that can be
expressed as a latent Gaussian Markov random files \citep[GMRF,][]{RueHeld:2005}.

In this context, the model is expressed as follows. For each observation
$y_i$ in the ensemble of observations $y$, its likelihood is defined as

$$
y_i |\bm\theta,\mathbf{x} \sim f(y_i|\bm\theta, \mathbf{x}),\ i=1,\ldots, n
$$

For each observation $y_i$, its mean $\mu_i$ will be conveniently linked to
the  linear predictor $\eta_i$ using the appropriate function (which will
depend on the likelihood used). The linear predictor will include different
additive effects such as fixed effects and random effects.

Next, the vector of latent effects $\mathbf{x}$ is defined as a GMRF with zero
mean and precision matrix $\Sigma$ (which may depend on the vector of
hyperparameters $\bm\theta$):

$$
\mathbf{x} \sim GMRF(0, \Sigma(\bm\theta))
$$

Finally, the hyperparameters are assigned a prior distribution. This is
often done assuming prior independence. Without loss of generality, this
can be represented as

$$
\bm\theta \sim \pi(\bm\theta)
$$
\noindent
Note that $\pi(\bm\theta)$ is a multivariate distribution but that in many
cases it can be expressed as the product of several univariate (or small
dimension) prior distributions.

Hence, if $\bm\theta$ represents the vector of $h$ hyperparameters and $\bm{x}$
the vector of $l$ latent effects, INLA provides the posterior marginals

$$
\{\pi(\theta_i | \mathcal{D})\}_{i=1}^{h}
$$
\noindent
and
$$
\{\pi(x_i | \mathcal{D})\}_{i=1}^{l}
$$
\noindent
Note that $\mathcal{D}$ represents the observed data, and it includes response
$y$, covariates $\mathbf{X}$ and any other known quantities required for model
fitting.

In addition to the marginals, INLA can be used to estimate other quantities
of interest. See, for example, \citet{INLAreview:2017} for a recent review.

INLA can provide accurate approximations to the marginal likelihood of a model,
which are computed as

$$
\tilde{\pi}(\mathcal{D}) = \int \frac{\pi(\bm{\theta}, \mathbf{x},
    \mathcal{D})}
{\tilde{\pi}_{\mathrm{G}}(\mathbf{x}|\bm{\theta},\mathcal{D})}
\bigg\lvert_{\mathbf{x}=\mathbf{x^*}(\bm\theta)}d \bm{\theta}.
$$
\noindent
Here, $\tilde{\pi}_{\mathrm{G}}(\mathbf{x}|\bm{\theta},\mathcal{D})$ is a
Gaussian approximation to the distribution of $\mathbf{x} \mid
\bm{\theta},\mathcal{D}$, and $\mathbf{x^*}(\bm\theta)$ is the posterior mode of
$\mathbf{x}$ for a given value of $\bm{\theta}$. This approximation
seems to be accurate in a wide range of examples \citep{HubinStorvik:2016b,GomezRubioRue:2018}.

As described in Section~\ref{sec:BMA}, the marginal likelihood plays a crucial
role in BMA as it determines the weights, together with the prior distribution
of some of the hyperparameters in the model.

As stated above, INLA approximates the posterior marginals of the parameters of
latent GMRF models. Hence, an immediate question is whether INLA will work for
different models. \cite{GomezRubioRue:2018} introduce the idea of using INLA to
fit \textit{conditional} latent GMRF models by conditioning on some of the
hyperparameters.  In this context, the vector of hyperparameters $\bm\theta$ is
split into $\bm\theta_c$ and $\bm\theta_{-c}$, so that models are fit with INLA
conditional on $\bm\theta_c$ and the posterior marginals of the elements of
$\bm\theta_{-c}$ and latent effects $\mathbf{x}$ are obtained, i.e.,
$\pi(\theta_{-c,i}\mid \mathcal{D}, \bm\theta_c)$ and $\pi(x_j \mid
\mathcal{D}, \bm\theta_c)$, respectively.
Here, $\theta_{-c,i}$ represents any element in $\bm\theta_{-c}$ and
$x_j$ any element in $\mathbf{x}$.

In practice, this involves setting hyperparameters $\bm\theta_c$ to
some value, so that the model becomes a latent GMRF that INLA can tackle.
\cite{GomezRubioRue:2018} provide some ideas on how these can be chosen.
\cite{GomezRubioPalmiPerales:2019} propose setting these values to maximum
likelihood estimates (for example) and other options and provide examples that
show that this may still provide good approximations to the posterior marginal
distributions of the remainder of the parameters in the model.

\section{Bayesian model averaging with INLA}
\label{sec:BMA}

As stated above, fitting conditional models by setting some hyperparameters
($\bm\theta_c$) to fixed values can be a way to use INLA to fit wider classes of
models. However, this ignores the uncertainty about these hyperparameters
$\bm\theta_c$ and makes inference about them impossible. However, BMA 
may help to fit the complete model, even if it cannot be fit with INLA
initially.

First of all, it is worth noting that the posterior marginals (of the
hyperparameters $\bm\theta_{-c}$ and latent effects $\mathbf{x}$) can
be written as

$$
\pi(\cdot \mid \mathcal{D}) = 
  \int \pi(\cdot, \bm\theta_{c} \mid \mathcal{D}) d\bm\theta_{c}= 
  \int \pi(\cdot  \mid \mathcal{D}, \bm\theta_{c}) 
    \pi(\bm\theta_{c} \mid \mathcal{D}) d\bm\theta_{c} 
$$
\noindent
The first term in the integrand, $\int \pi(\cdot  \mid \mathcal{D}, \bm\theta_{c})$,
 is the conditional posterior marginals given $\bm\theta_{c}$, while
the second term is the joint posterior distribution of $\bm\theta_{c}$ and it
 can be expressed as

$$
\pi(\bm\theta_{c} \mid \mathcal{D}) \propto \pi(\mathcal{D} \mid \bm\theta_{c}) \pi(\bm\theta_{c})
$$
\noindent
The first term is the conditional (on $\bm\theta_{c}$) marginal likelihood
which can be approximated with INLA. The second term is the prior for
$\bm\theta_{c}$, which is known. Hence, the posterior distribution of
$\bm\theta_{c}$ could be computed by re-scaling the previous expression.

\cite{Bivandetal:2014} show that when $\bm\theta_{c}$ is unidimensional,
numerical integration can be used to estimate the posterior marginal. This is
done by using a regular grid of $K$ values $\{\theta^{(i)_c}\}_{i=1}^K$ of
fixed step $h$.

Hence, the posterior marginals of the remainder of hyperparameters and
latent effects can be estimated as

$$
\pi(\cdot \mid \mathcal{D}) \simeq
\sum_{i=1}^K \pi(\cdot \mid \mathcal{D}, \theta_{c}^{(i)})\ w_i
$$
\noindent
with weights $w_i$ defined as

$$
w_i = \frac{
  \pi(\mathcal{D} \mid \theta^{(i)}_{c}) \pi(\theta_{c}^{(i)})}
  {\sum_{i=1}^K \pi(\mathcal{D} \mid \theta^{(i)}_{c}) \pi(\theta_{c}^{(i)}) } 
$$

Note how the posterior marginal $\pi(\cdot \mid \mathcal{D})$ is expressed
as a BMA using the conditional posterior marginals of all the fit models.

Inference on $\bm\theta_c$ is based on the values $\{\theta^{(i)_c}\}_{i=1}^K$
and weights $\{ w_i \}^K$. For example, the posterior mean of the 
element $j$-th in $\bm\theta_c$ could be computed as 
$\sum_{i=1}^K \theta_{c,j} w_i$. Other posterior quantities could be computed
similarly. Note that this also allows for multivariate posterior inference
on the elements of $\bm\theta_c$.

This can be easily extended to higher dimensions by considering 
that the proposed values of $\bm\theta_c$ fill the space in subsets of
equal volume. In practice, it is not necessary to consider the complete
space (as it is not feasible) but the region of posterior high probability
of $\bm\theta_c$. This may be obtained by, for example, maximizing
$\pi(\mathcal{D} \mid \bm\theta_{c}) \pi(\bm\theta_{c})$, which can be easily
computed with INLA or by using a maximum likelihood estimate if available
\citep[as discussed, for example, in][]{GomezRubioPalmiPerales:2019}.

\section{Example: turnover in Italy} \label{sec:example}

In order to provide an example of the methodology presented in previous
sections we will take the turnover dataset described in
\cite{WardGledistch:2008}, which is available from website
\url{http://ksgleditsch.com/srm_book.html}. This dataset records the turnover
in the 2001 Italian elections, as well as GDP per capita (\texttt{GDPCAP}) in
1997 (in million Lire).  The data comprises 477 areas, which represent
\textit{collegi} or single member districts (SDM). Adjacency has been defined
so that regions whose centroids are 50 km or less away are neighbors to ensure
that all regions have neighbors, contiguous regions are neighbors, and Elba is
joined to the closest mainland region.  In order to assess the impact of the
GDP per capita in the estimation of the spatial autocorrelation parameters, two
models with and without  the covariate will be fit.

\begin{figure}[h!]
\includegraphics[height = 8cm]{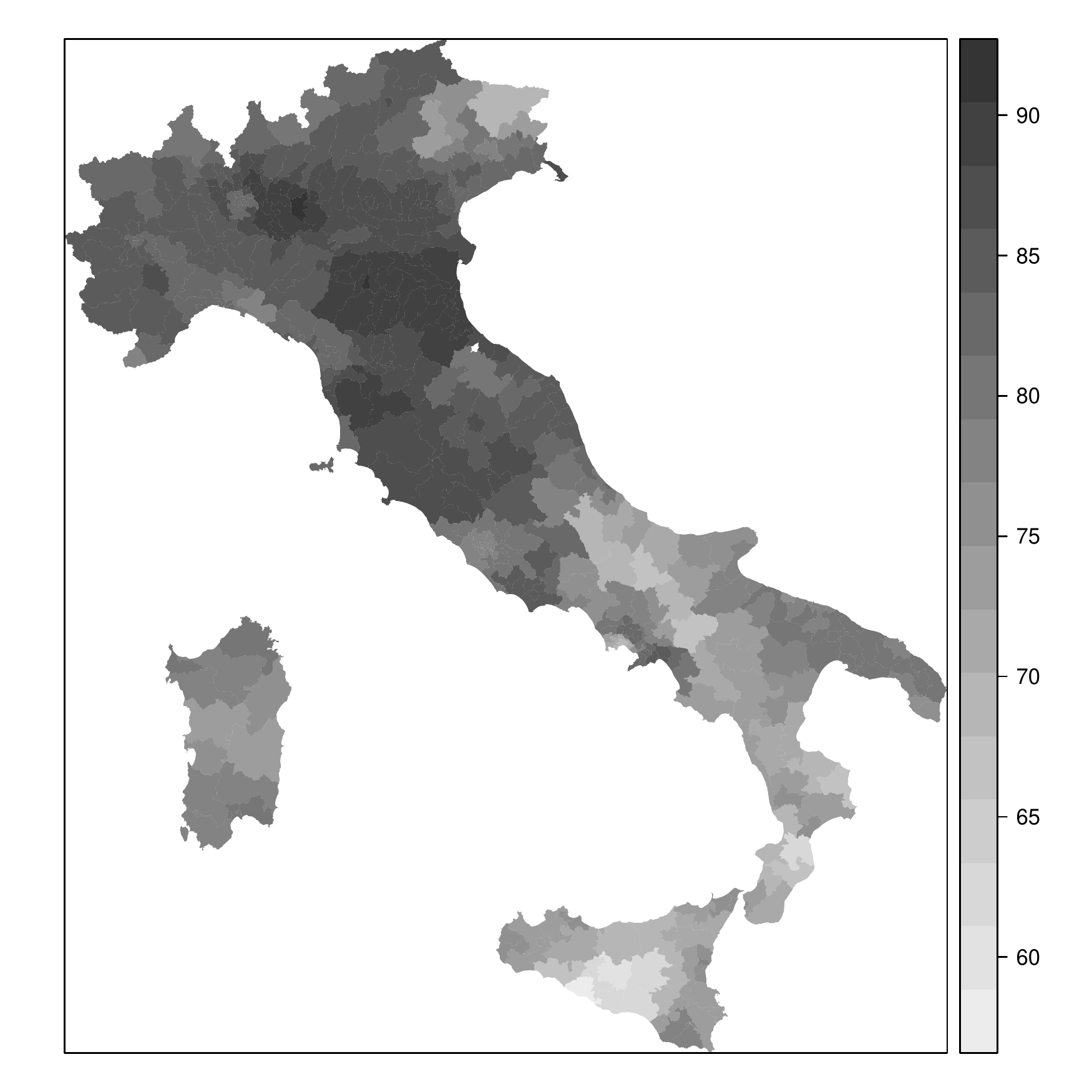}
\includegraphics[height = 8cm]{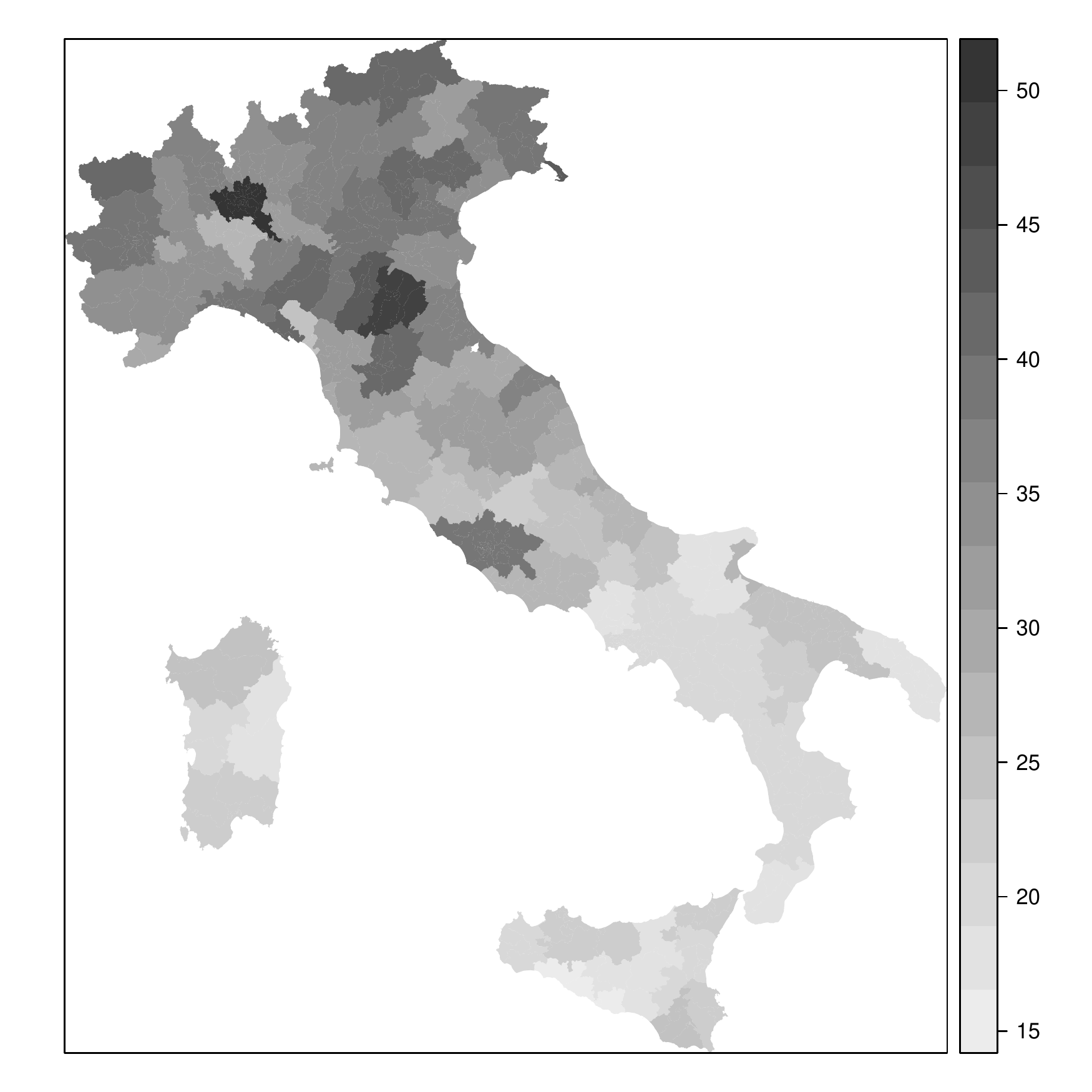}

\caption{Spatial distribution of turnover in 2011 (left) and GDP per capita in 1997 (right) in Italy at the \textit{collegi} level.}
\label{fig:maps}
\end{figure}

Figure~\ref{fig:maps} shows the spatial patterns of these two variables.  As it
can be seen, there is a clear south-north pattern for both variables.  Hence,
we are interested in fitting spatial econometrics models on turnover in 2011
using GDP per capita in 1997 as predictor so that residual spatial
autocorrelation is captured by the two autoregressive terms in the model.

The SAC model will be fit with INLA by conditioning on the values of the two
spatial autocorrelation parameters. As described in Section~\ref{sec:SE}, after
conditioning the resulting model is a typical mixed-effects model with a
particular matrix of covariates and a known structure for the precision matrix
of the random effects. Hence, we are considering $\bm\theta_{c} = (\rho,
\lambda)$.

Both autocorrelation parameters are given values in the range (-1, 1).  Note
that, because the same adjacency matrix is used, the actual domain for both
parameters is $(1 / \lambda_{min}, 1)$, with $\lambda_{min}$ the minimum eigenvalue
of the adjacency matrix $W$. In this case, $\lambda_{min} = -0.82$ so  taking
the range $(-1, 1)$ will be enough.

Maximum likelihood estimates of the models will be obtained with package
\texttt{spdep} \citep{Bivandetal:2013}, and MCMC estimates will be obtained
with package \texttt{spatialreg} \citep{BivandPiras:2015,Bivandetal:2013}.  For
MCMC, we used 10000 burn-in iterations, plus another 90000 iterations for
inference, of which only one in ten was kept to reduce autocorrelation.  To
speed up convergence, the initial values of $\rho$ and $\lambda$ were set to
their maximum likelihood estimates. BMA with INLA estimates will be obtained as
explained next.

%A regular grid to about the posterior mode of $(\rho, \lambda)$ will be
%created and this will different for both models. For the model with the
%intercept only the region explored is $[0.88, 0.90]\times [0.29, 0.31]$, while
%for the model with covariates it is $[0.82, 0.94]\times [0.465, 0.485]$. The
%step is $0.0015$ for both parameters.

A regular grid about the posterior mode of $(\rho, \lambda)$ will be created
for each model using the maximum likelihood estimates.  The grid is defined in
an internal scale to have unbounded variables using the transformation
$\gamma_1 = \log(\frac{1 + \rho}{1 - \rho})$ and $\gamma_2 = \log(\frac{1 +
\lambda}{1 - \lambda})$. The variance of $\gamma_1$ and $\gamma_2$ can be
derived using Delta method (see Appendix \ref{app} for details), and it can be
computed  using the ML standard error estimate of $\rho$ and $\lambda$.  In
particular, each interval is centered at the transformed ML estimate and a
semi-amplitude of three standard errors of the variables in the internal scale.
See Table~\ref{tab:params} for the actual values of the ML estimates.
Furthermore, a grid of $160 \times 40$ points was used to represent the search
space and fit the conditional models with R-INLA for the model without the
covariate and a grid of $40 \times 20$ for the model with \texttt{GDPCAP}. A
different grid was used because the model without the covariate required a
thinner grid.

Table~\ref{tab:params} provides a summary of the estimates of the model
parameters using different inference methods. First of all, maximum likelihood
(ML) estimates (computed with function \texttt{sacsarlm} in the \texttt{spdep}
package). Next, $\rho$ and $\lambda$ have been fixed at their ML estimates and
the model has been fit with INLA. Next, the  posterior marginals of the model
parameters using BMA with INLA and MCMC are shown. In general, point estimates
obtained with the different methods provide very similar values. MCMC and
maximum likelihood also provide very similar estimates of the uncertainty of
the point estimates (when available for maximum likelihood).
BMA with INLA seems to provide very similar results to MCMC for both models.

\begin{table}[h!]
\centering
\begin{tabular}{c|c|c|c||c|c||c|c||c|c}
 &  & \multicolumn{2}{c||}{Max. lik.} & \multicolumn{2}{c||}{INLA - Max. lik.}  &
   \multicolumn{2}{c||}{BMA} & \multicolumn{2}{c}{MCMC}\\
 \cline{3-10} 
Model & Parameter & Mean & St. error  & Mean & St. dev. & Mean & St. dev. & 
   Mean & St. error \\
\hline
\multirow{3}{*}{No covariates} &
   $\beta_0$ & 5.86 & 1.54 & 5.88 & 0.10 & 6.39 & 1.59  & 6.56 & 1.95\\
 & $\rho$ & 0.93 & 0.02 & 0.93 & -- & 0.92 & 0.02 & 0.92 & 0.02\\
 & $\lambda$ & 0.09 & 0.10 & 0.09 & -- & 0.12 & 0.10 & 0.12 & 0.10\\
 & $\tau^{-1}$ & 3.66 & -- & 3.68 & 0.24 & 3.71 & 0.25 & 3.75 & 0.28 \\
\hline
\hline
 \multirow{4}{*}{Covariates} &
   $\beta_0$ & 8.33 & 2.46 & 8.33 & 0.37 & 9.46 & 2.44 & 9.53 & 3.04\\
 & $\beta_1$ & 0.05 & 0.02 & 0.05 & 0.01 & 0.05 & 0.02 & 0.05 & 0.02 \\
 & $\rho$ & 0.88 & 0.03 & 0.88 & -- & 0.86 & 0.03 & 0.86 & 0.04\\
 & $\lambda$ & 0.17 & 0.11 & 0.17 & -- & 0.22 & 0.11 & 0.21 & 0.12\\
 & $\tau^{-1}$ & 3.77 & -- & 3.79 & 0.24 & 3.82 & 0.26 & 3.89 & 0.31\\
\hline
\end{tabular}
\caption{Summary statistics of the model parameters.}
\label{tab:params}
\end{table}

Note the positive coefficient of GDP per capita, which means a positive
association with turnout. Furthermore, the spatial autocorrelation $\rho$
has a higher value than $\lambda$, which indicates higher autocorrelation
on the response than on the error term.

\begin{figure}[h]
\centering
\includegraphics[width=12cm]{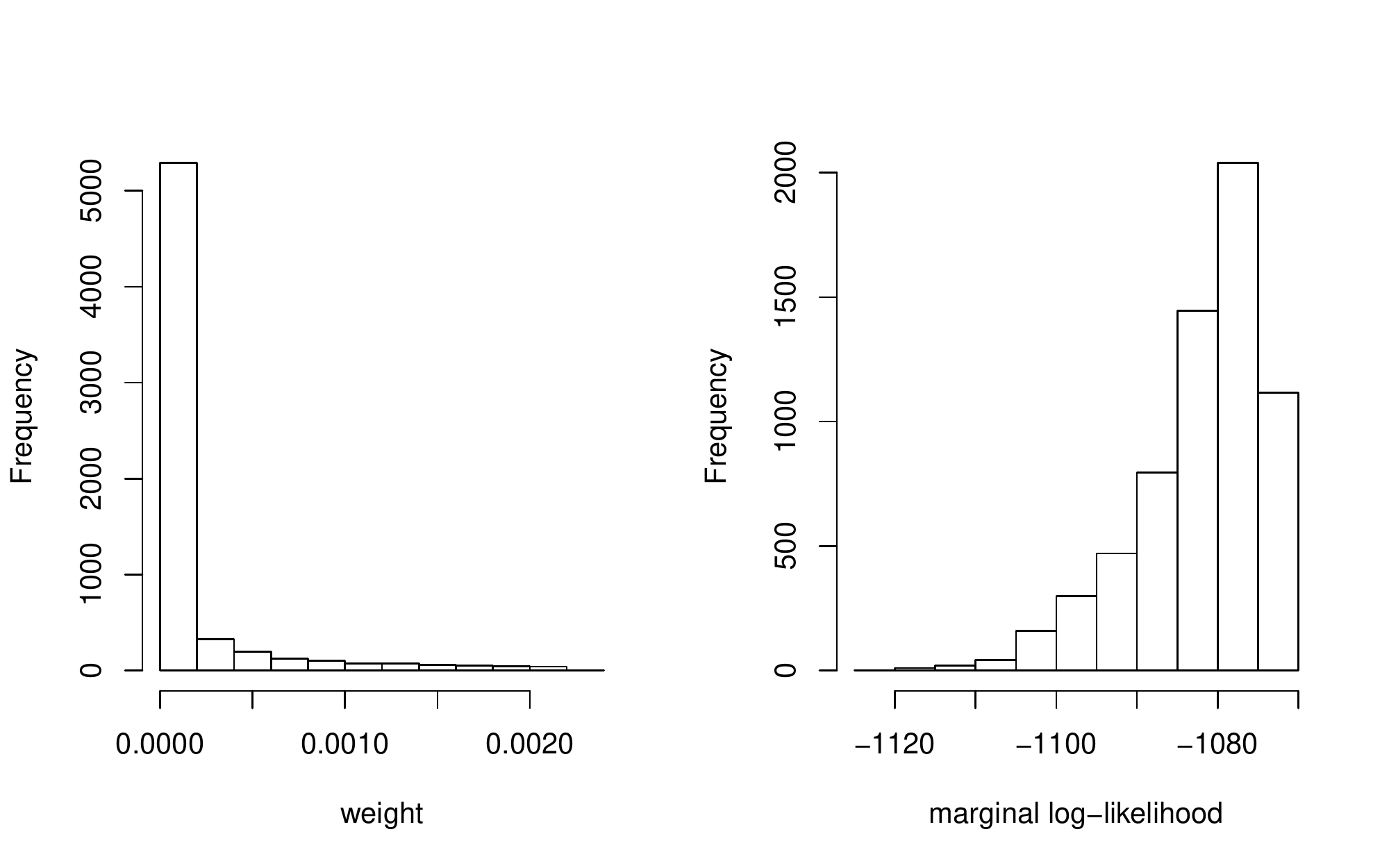}
\includegraphics[width=12cm]{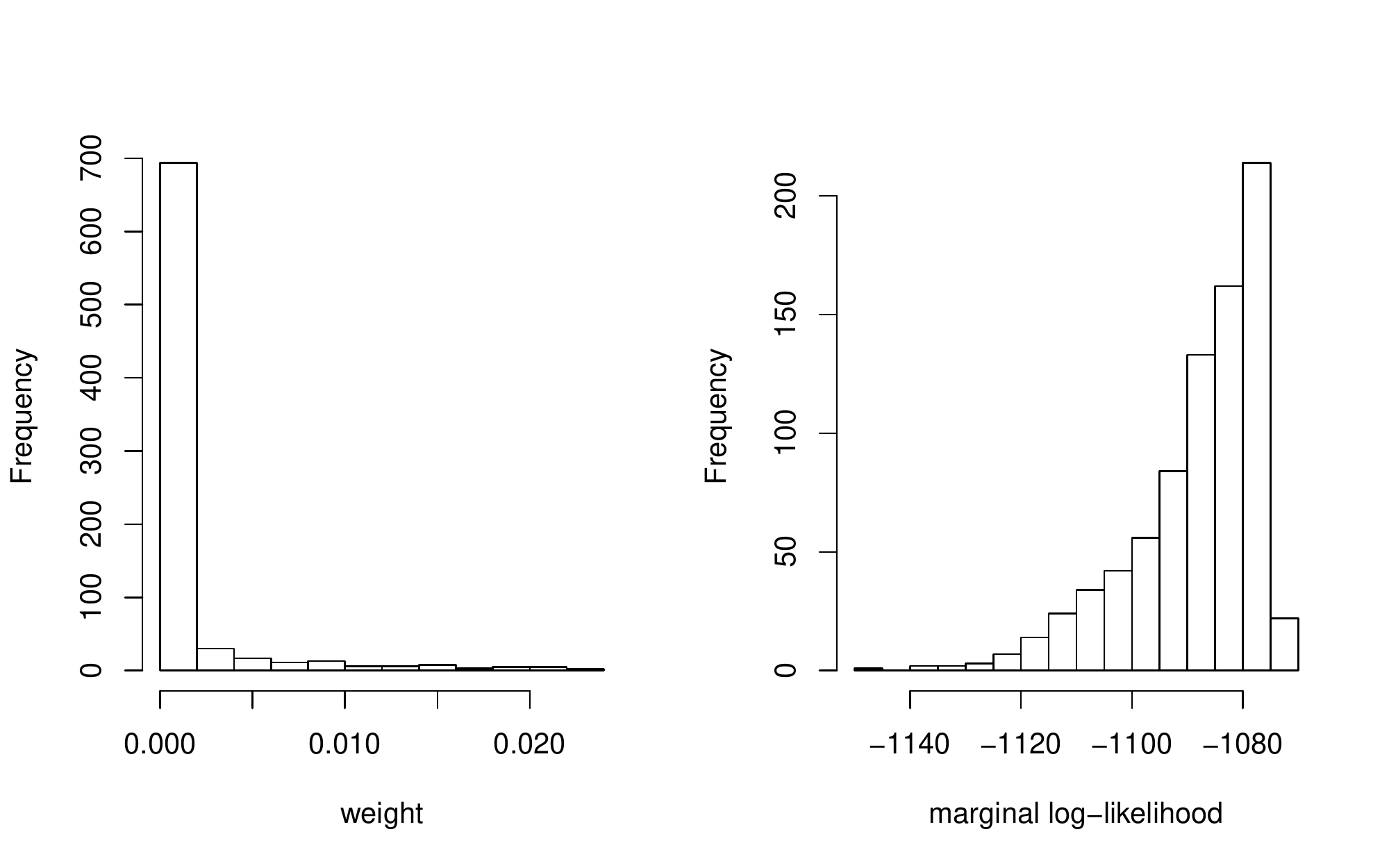}
\caption{Weights and marginal likelihoods for the model with no covariates (top row) and with covariates (bottom row).}
\label{fig:weights}
\end{figure}

Figure~\ref{fig:weights} shows the values of the marginal log-likelihoods of
all the conditional models fit as well as the weights used in BMA. Note the
wide variability of the marginal log-likelihoods that makes the weights to
be very small. Weights are almost zero for most models but a few of them.
Note that this is good because it indicates that the search space for
$\rho$ and $\lambda$ is adequate and that the region of high posterior
density has been explored.

\begin{figure}[h]
\centering
\includegraphics[width=12cm]{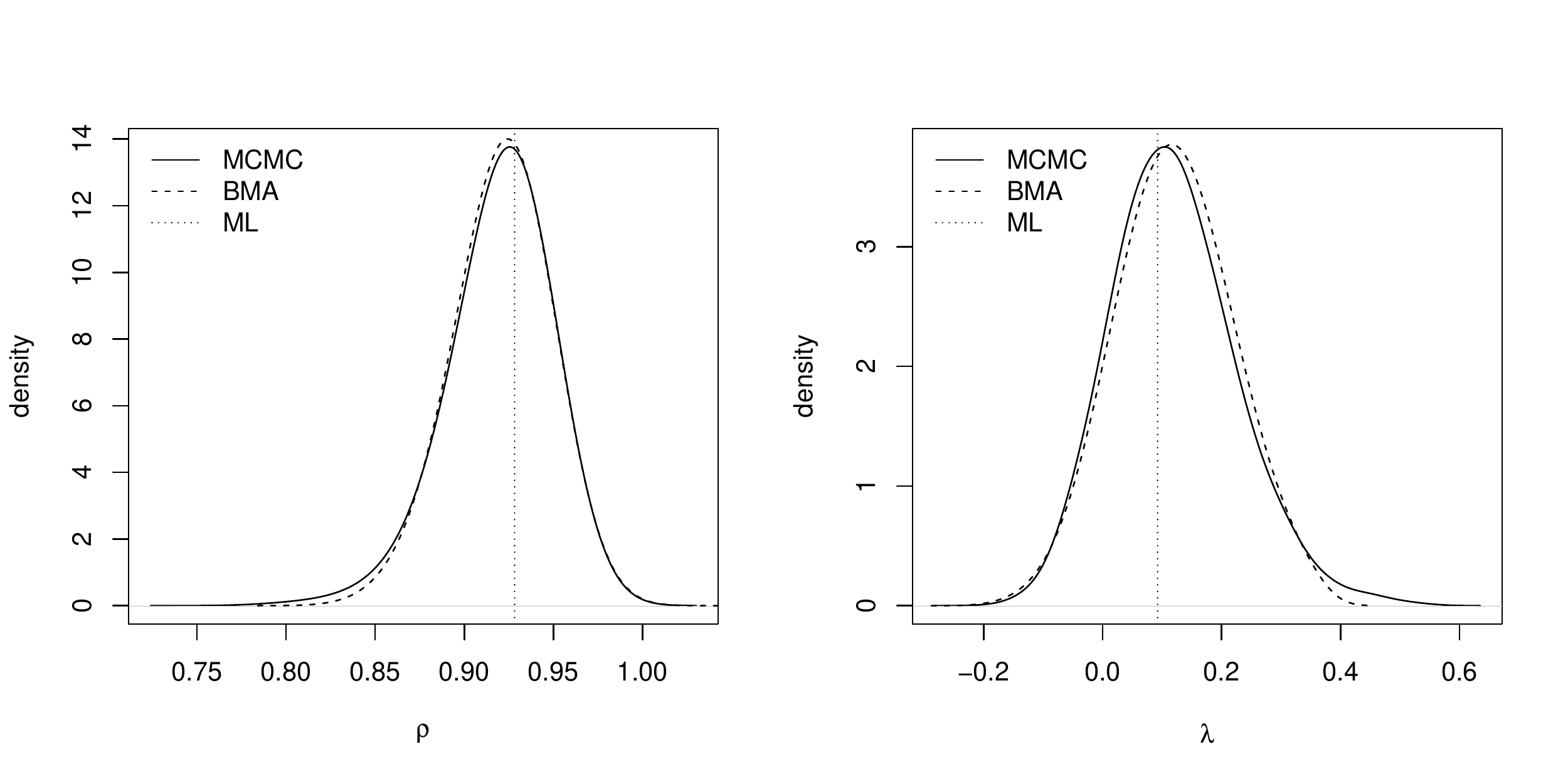}
\includegraphics[width=12cm]{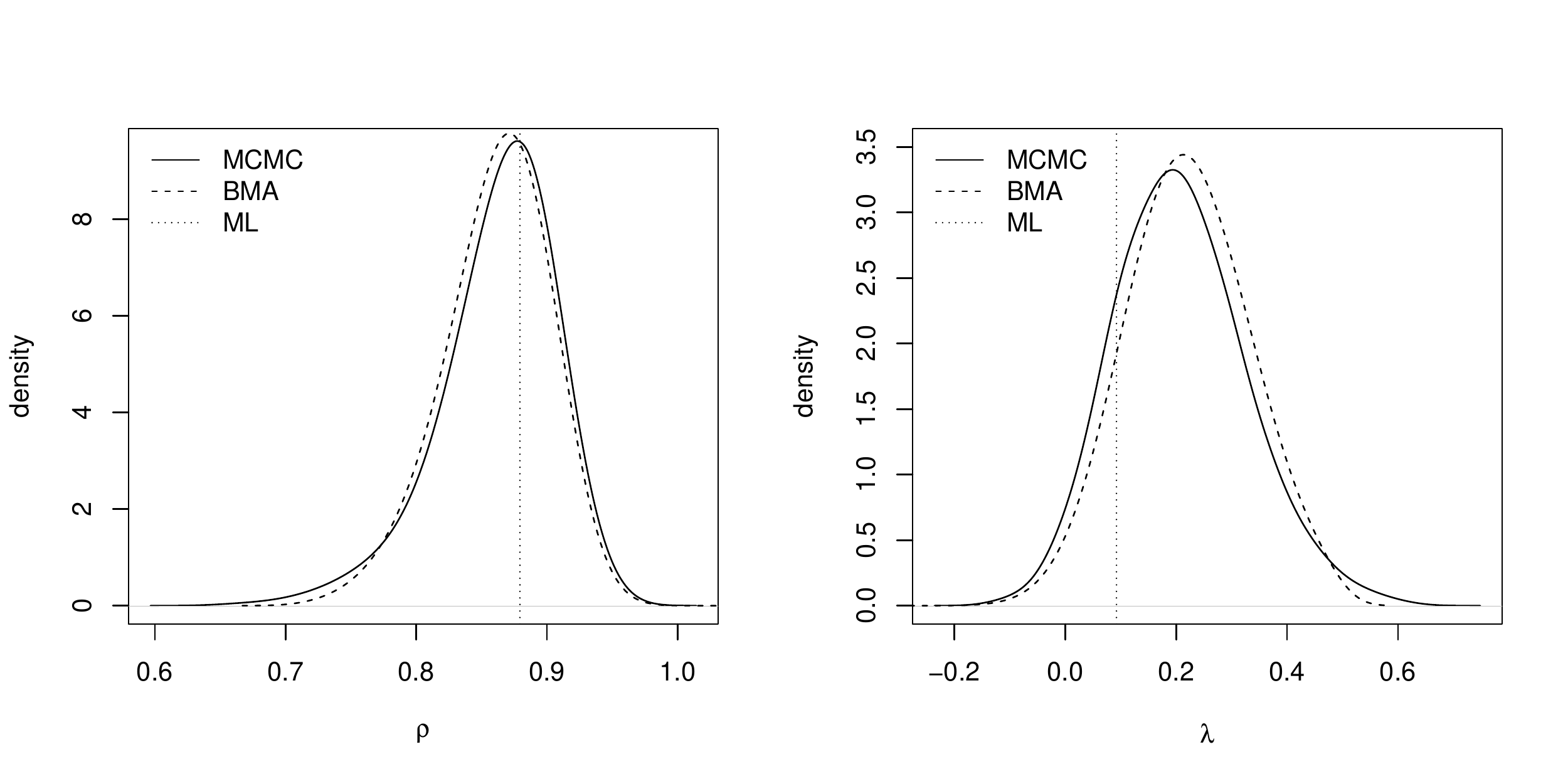}
\caption{Posterior marginals of the spatial autocorrelation parameters for the model with no covariates (top row) and with covariates (bottom row).}
\label{fig:margs}
\end{figure}

Figure~\ref{fig:margs} shows the posterior marginals of the spatial
autocorrelation parameters using kernel smoothing \citep{MASS}. For the BMA
output weighted kernel smoothing, has been used. In general, there is good
agreement between BMA with INLA and MCMC. The ML estimates are also very close
to the posterior modes.

\begin{figure}[h]
\centering
\includegraphics[width=12cm]{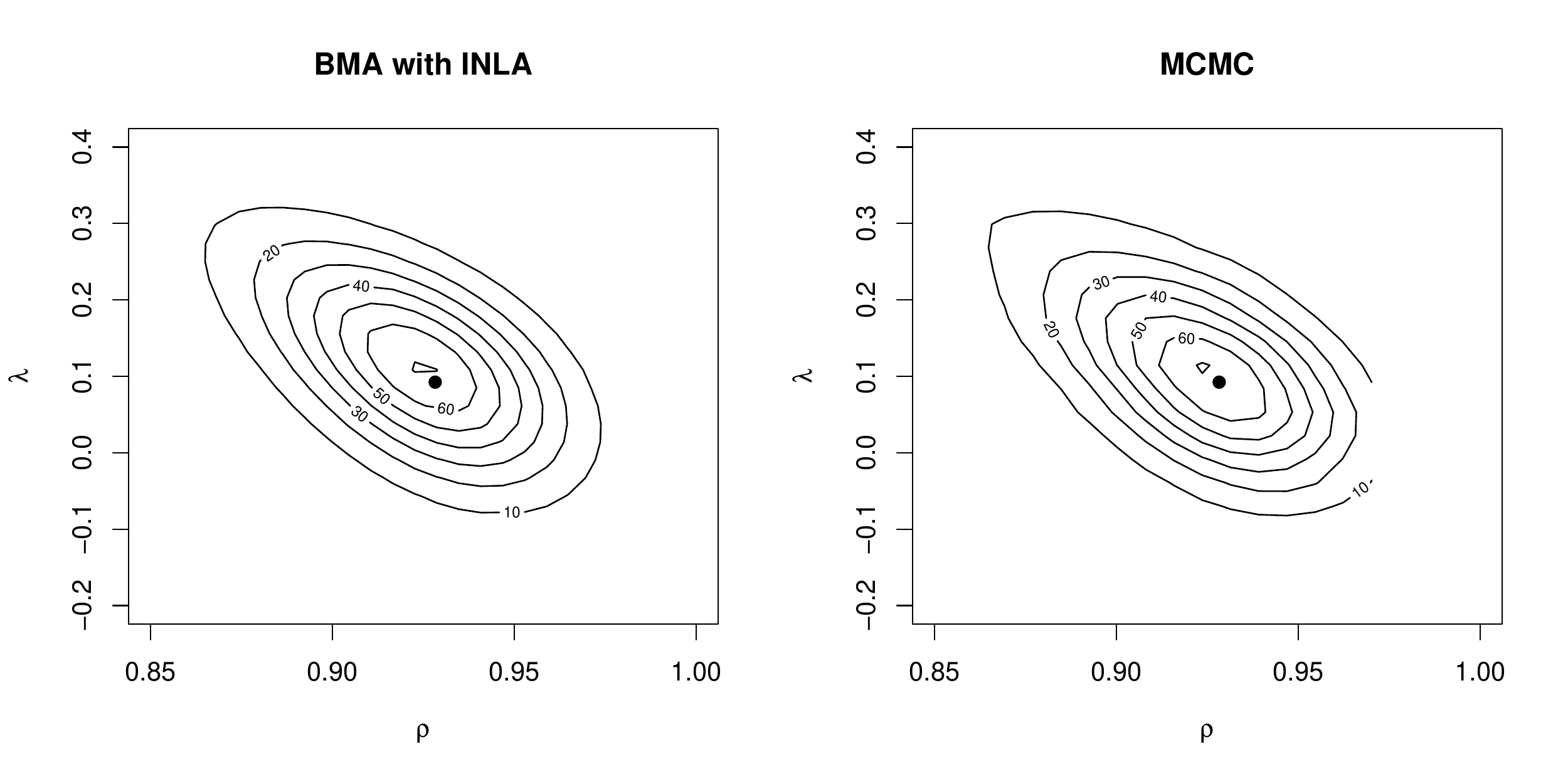}
\includegraphics[width=12cm]{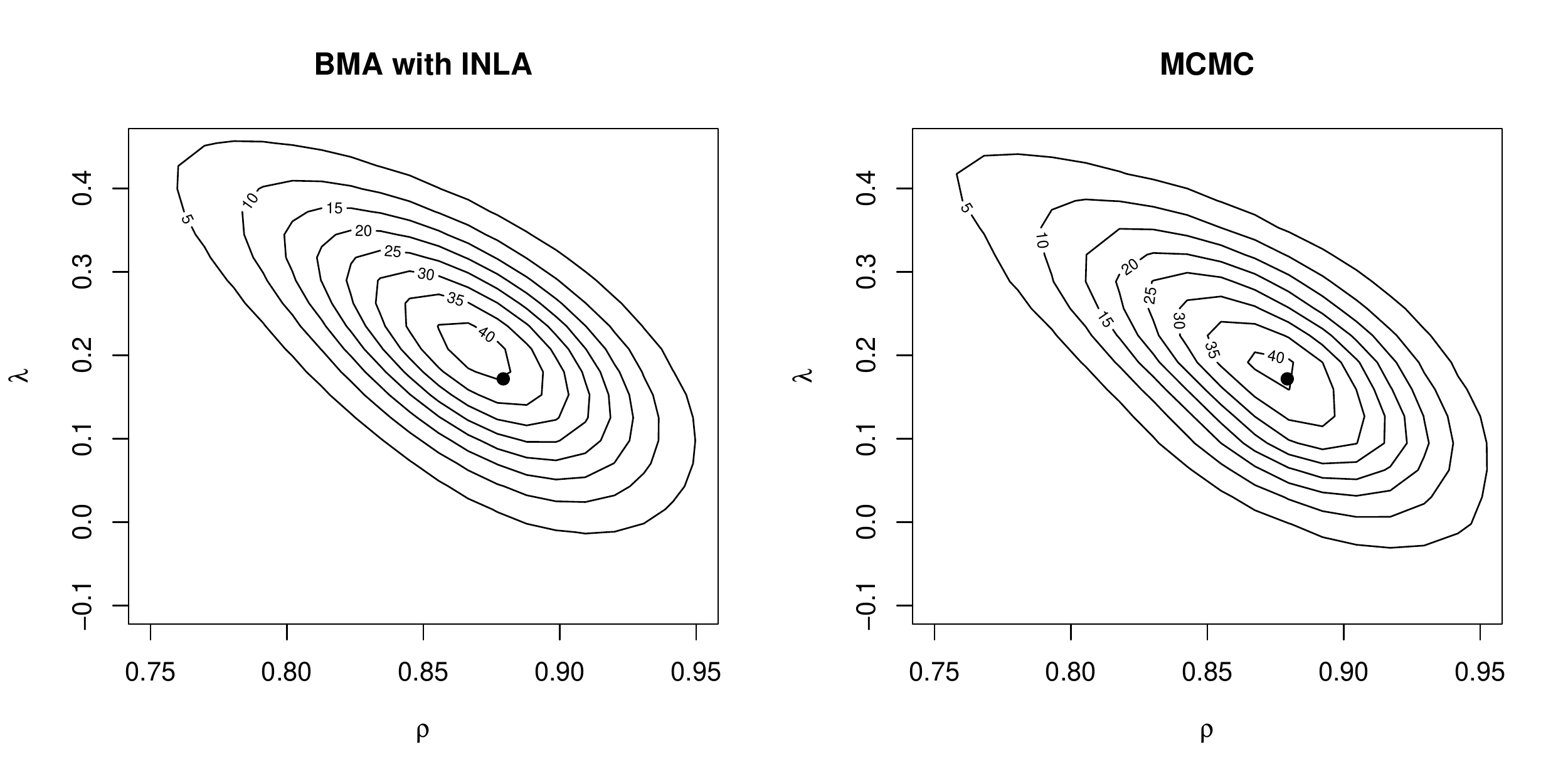}
\caption{Joint posterior distribution of the spatial autocorrelation parameters for the model with no covariates (top row) and with covariates (bottom row). The black dot represents the maximum likelihood estimate.}
\label{fig:biplots}
\end{figure}

Similarly, Figure~\ref{fig:biplots} shows the joint posterior distribution of
the autocorrelation parameters obtained using two-dimensional kernel smoothing.
For the BMA output this has been obtained using two-dimensional weighted kernel
smoothing with function \texttt{kde2d.weighted} in package \texttt{ggtern}
\citep{ggtern}. The ML estimate has also been added (as a black dot). As with
the posterior marginals, the joint distribution is close between MCMC and BMA
with INLA.  The posterior mode is also close to the ML estimate.  The plots
show a negative correlation between the spatial autocorrelation parameters,
which may indicate that they struggle to explain spatial correlation in the
data \citep[see also ][]{GomezRubioPalmiPerales:2019}.
Furthermore, BMA with INLA is a valid approach to make joint posterior 
inference on a subset of hyperparameters in the model.

These results shows the validity of relying on BMA with INLA to fit highly
parameterized models. This approach also  accounts for the uncertainty of all
model parameters and should be preferred to other inference methods based on
plugging-in or fixing some of the models parameters. In our case, we have
relied on BMA with INLA so that conditional sub-models are fit and then
combined. This has two main benefits. First of all, allows for full inference
on all model parameters and, secondly, uncertainty about all parameters is
taken into account.

\begin{figure}[h]
\centering
\includegraphics[width=12cm]{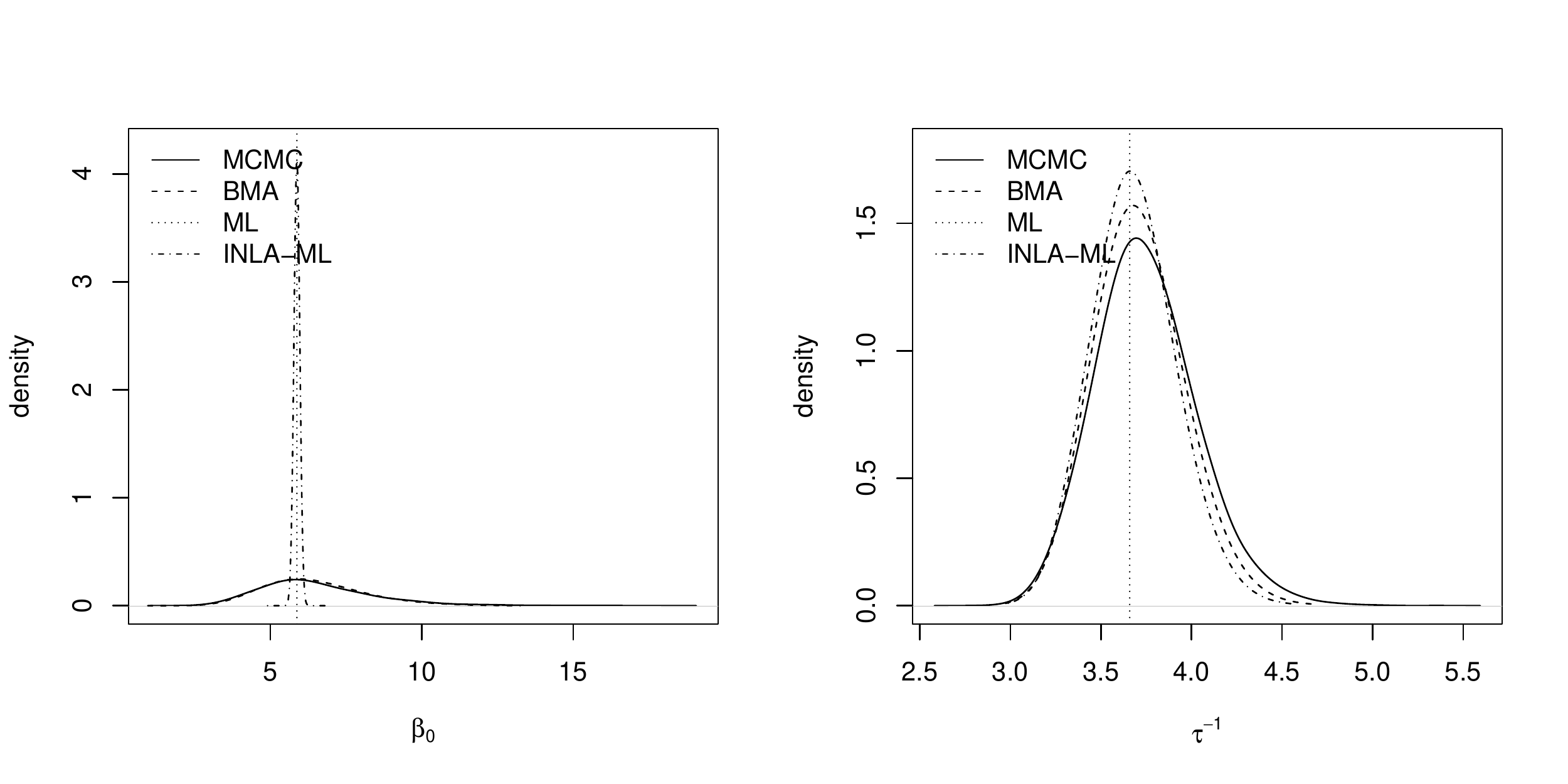}
\includegraphics[width=12cm]{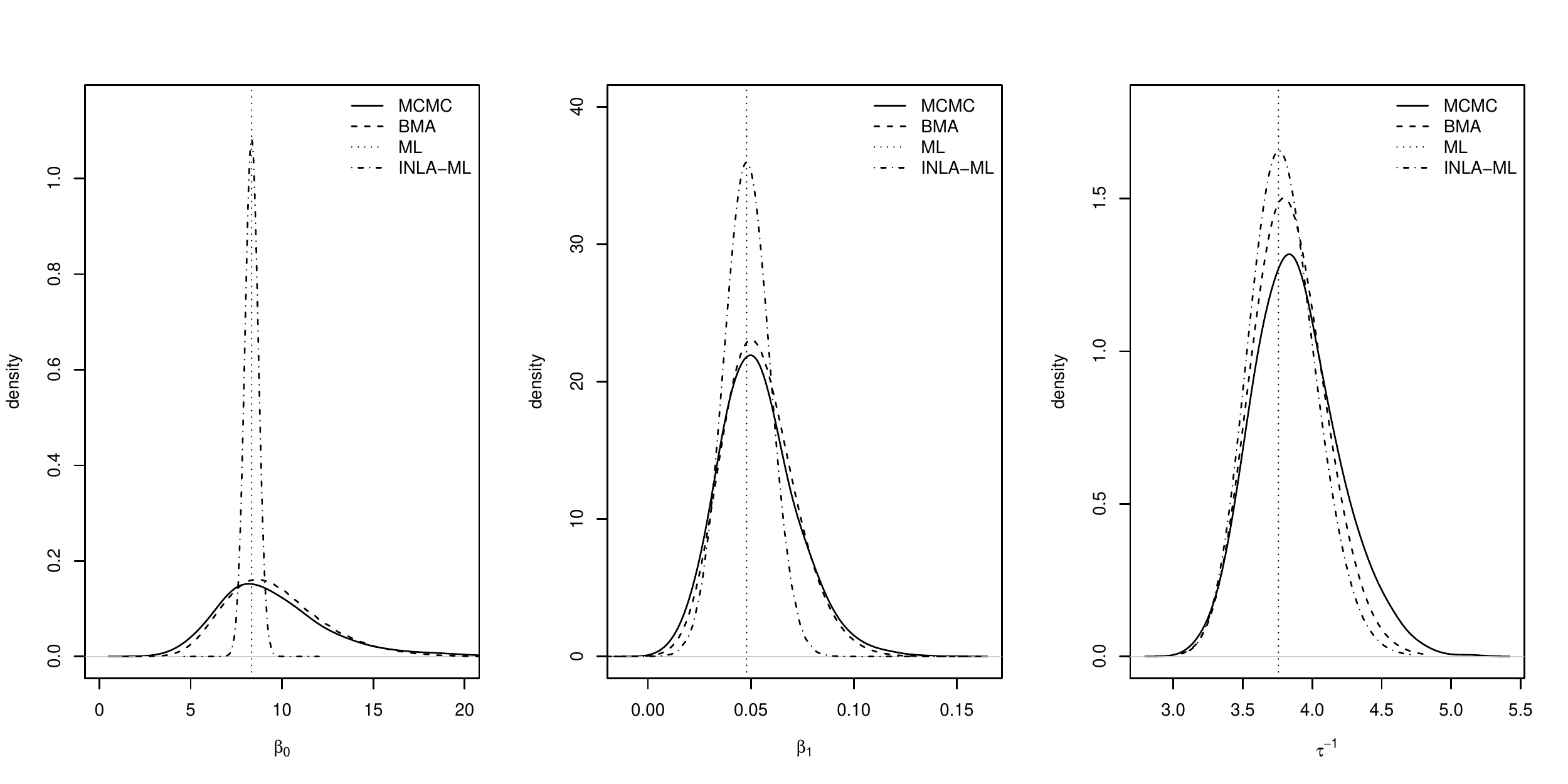}
\caption{Posterior marginal distributions of the coefficients and variances  for the model with no covariates (top row) and with covariates (bottom row).}
\label{fig:params}
\end{figure}

Figure~\ref{fig:params} shows the posterior marginals of the coefficients and
the variance of the models. In general, BMA with INLA and MCMC show very
similar results for all parameters The posterior modes of MCMC and BMA with
INLA are very close to the ML estimates.  Marginals provided by INLA with ML
estimates are very narrow for the fixed effects, which is probably due to
ignoring the uncertainty about the spatial autocorrelation parameters.

\begin{figure}[h]
\centering
\includegraphics[width=12cm]{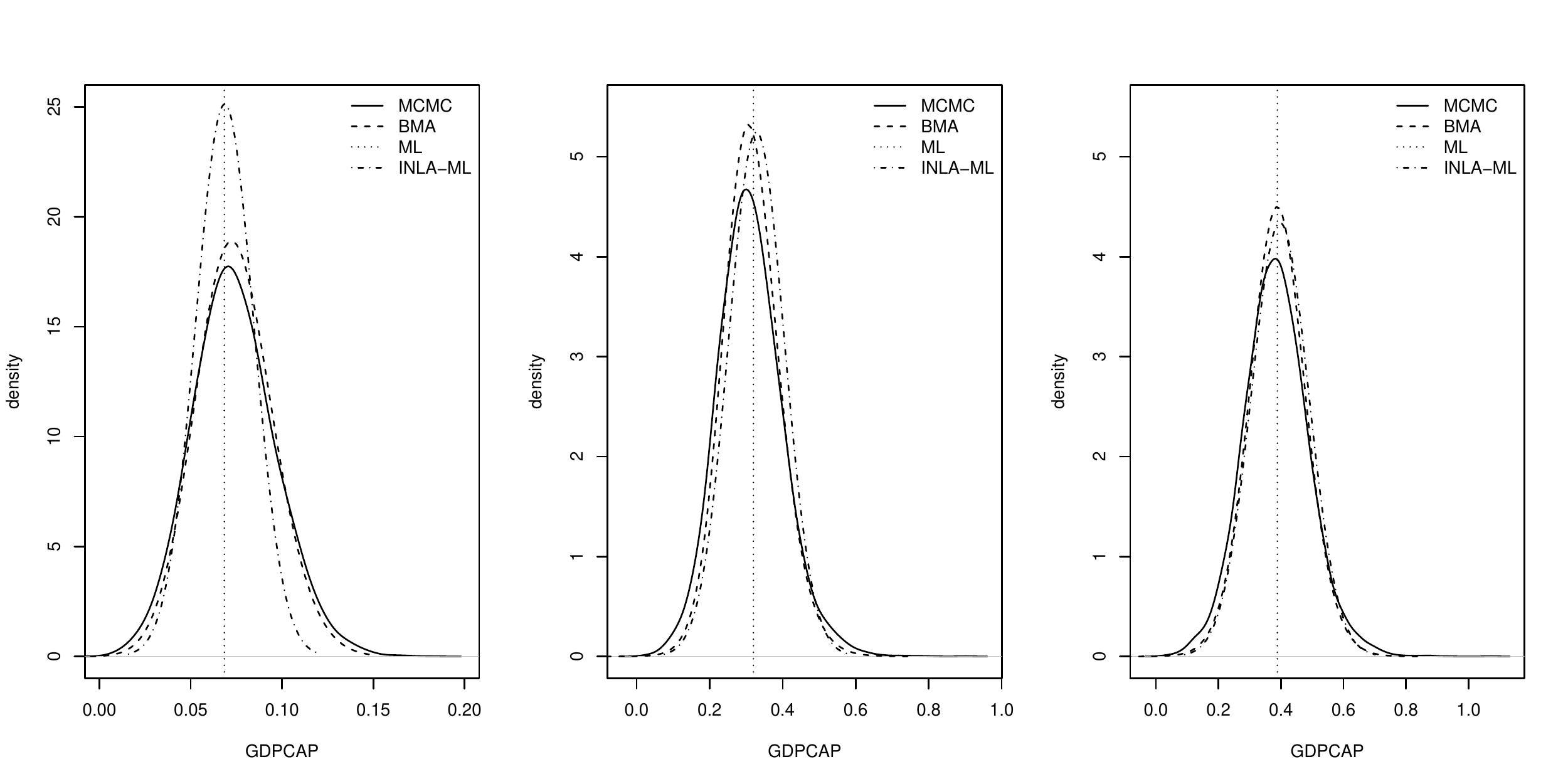}
\caption{Posterior marginals of the average direct (left), indirect (middle) and total (right) impacts of the GDP per capita.}
\label{fig:impacts}
\end{figure}

Computation of the impacts is important because they measure how changes in the
values of the covariates reflect on changes of the response variable in the
current area (direct impacts) and neighboring areas (indirect impacts). Note
that impacts are only computed for the model with the covariate included.
Table~\ref{tab:impacts} shows the estimates of the different average impacts.
In general, all estimation methods provide very similar values of the 
point estimates, and BMA with INLA and MCMC estimates are very close.
Figure~\ref{fig:impacts} displays the posterior marginal distributions of the
average impacts. Again, these estimates are very similar among estimation
methods.

\begin{table}[h!]
\begin{tabular}{c|c|c||c|c||c|c||c|c}
& \multicolumn{2}{c||}{Max. lik.} & \multicolumn{2}{c||}{INLA - Max. lik.} & \multicolumn{2}{c||}{BMA} & \multicolumn{2}{c}{MCMC} \\
\cline{2-9}
Impact & Mean & St. dev. &Mean & St. dev. &Mean & St. dev. &Mean & St. dev.\\
\hline
Direct &   0.07 & -- & 0.07 & 0.02 & 0.07 & 0.02 & 0.07 & 0.02 \\
Indirect & 0.32 & -- & 0.33 & 0.08 & 0.32 & 0.08 & 0.31 & 0.09 \\
Total &    0.39 & -- & 0.40 & 0.09 & 0.39 & 0.09 & 0.38 & 0.10 \\
\hline
\end{tabular}
\caption{Average impacts estimated with the different methods.}
\label{tab:impacts}
\end{table}

\section{Discussion}
\label{sec:discussion}

Bayesian model averaging with the integrated nested Laplace approximation has
been illustrated to make inference about the parameters and latent
effects of highly parameterized models. The appeal of this methodology is that
it relaxes the constraints on the model that does not need to be latent GMRF
anymore but \textit{conditional} latent GMRF. This has been laid out with an
example based on spatial econometrics models.

Cheaper alternatives to BMA include setting the values of a subset
of the hyperparameters to their posterior modes or maximum likelihood
estimates. This can still produce accurate estimates of the posterior
marginals of the remainder of the parameters in the model \citep{GomezRubioPalmiPerales:2019} but ignores the uncertainty about some parameters in the model
as well as any posterior inference about them.

Although we have used numerical integration methods to estimate the joint
posterior distribution of the conditioning hyperparameters, this is limited to
low dimensions and it may not scale well. However, other approaches could be
used such as MCMC algorithms \citep{GomezRubioRue:2018}.

Finally, BMA with INLA makes inference about a small subset of hyperparameters
in the mode possible. This is an interesting feature as INLA focuses on marginal
inference and joint posterior analysis require sampling from the internal
representation of the latent field \citep{GomezRubio:2019}, which can be costly.
We have also shown how this can be used to compute the posterior marginals
distribution of derived quantities (i.e., the impacts) that depend on a small
subset of hyperparameters. Our results with BMA with INLA are very close to
those obtained with MCMC, which supports the use of BMA with INLA as a
feasible alternative for multivariate posterior inference.

\appendix

\section{Implementation details}
\label{app}

\subsection{Model fitting}
Implementation of the SAC model with INLA is done in two steps. First,
the model is fit given $\rho$ and $\lambda$. Given these values, the model
is a linear mixed-effects model with design matrix of the fixed effects equal to

$$
(I - \rho W) ^{-1} X 
$$
\noindent
and the vector of random effects has mean zero and precision matrix

$$
\tau \Sigma,
$$
\noindent
where $\tau$ is a precision parameter to be estimated. Matrix $\Sigma$ is a
precision matrix fully determined (given that $\rho$ and $\lambda$ are known)
and it is 

$$
\Sigma = 
  (I - \rho W^{\top}) (I - \lambda W^{\top}) (I - \lambda W) (I - \rho W).
$$ 

This model can be fit with the R-INLA package using a Gaussian likelihood, and
a design matrix for the fixed effects $(I - \rho W) ^{-1} X$ and latent random
effects with zero mean and precision matrix $\tau \Sigma$. In particular, the
latent random effects can be defined using effect \texttt{generic0}.
Furthermore, the precision of the Gaussian likelihood is set to $\exp(15)$ to
remove the error term. Otherwise, the model would include another error term
in the linear predictor in addition to fixed and random effects. Higher
values of the precision led to unstable estimates of the model and
the marginal likelihood.

The \texttt{generic0} latent effects is a multivariate Gaussian distribution
with zero mean and precision matrix $\tau \Sigma$. Because $\Sigma$ is known,
its determinant is ignored by R-INLA when computing the log-likelihood. For
this reason, we have added $+\frac{1}{2}\log(|\Sigma|)$ to the computation of
the marginal likelihood reported by R-INLA, with $|\Sigma|$ the determinant of
$\Sigma$.

Once the different models have been fit, weights $w_i$ are computed as in
Section~\ref{sec:BMA}. Models are merged using function \texttt{inla.merge},
that takes the list of models fit and the vector of weights.

\subsection{Grid definition}

As stated in Section~\ref{sec:example} the grid to explore the values of
$(\rho, \lambda)$ is a regular grid defined in an \textit{internal} scale to
make the \textit{internal} parameters unbounded. The estimate of the variance
of the parameters in the internal scale can be obtained by using the Delta method.
The transformation used is $\gamma_1 = \log(\frac{1 + \rho}{1 - \rho})$ and
$\gamma_2 = \log(\frac{1 + \lambda}{1 - \lambda})$. If we consider $g(x) =
\log(\frac{1 + x}{1 - x})$, then $\gamma_1 = g(\rho)$ and $\gamma_2 =
g(\lambda)$.  Hence, the inverse transformation can be considered if we take
function $f(x) = 2 \frac{\exp(x)}{1 + \exp(x)} -1$, and the original parameters are
defined as $\rho = f(\gamma_1)$ and $\lambda = f(\gamma_2)$.

The Delta method states that if $\sigma^2$ is the variance of a variable $x$,
then an estimate of the variance of $g(x)$ is $\sigma^2 (g^{'}(x))^2$, with
$g^{'}(x)$ the first derivative of $g(x)$. 
In our case, we have that 

$$
g^{'}(x)  = \frac{2}{(1 + x) (1 - x)}.
$$

Using ML estimation, for both $\rho$ and $\lambda$
an estimate of the variance is provided by their respective squared standard 
errors. Hence, the estimate of the variance of $\gamma_1$ is 

$$
se_{\rho}^2 (\frac{2}{(1 + \rho) (1 - \rho)})^2
$$
\noindent
Here, $se_{\rho}$ is the standard error of $\rho$ and $\rho$ represents the ML
estimate. Similarly, an estimate of the variance of $\gamma_2$ is developed.

Note that because the regular grid is defined in the internal scale,
the prior should be on $\gamma_1$ and $\gamma_2$. This can be derived
\citep[Chapter 5, ][]{GomezRubio:2019} by considering a correction
due to the transformation. For example, the prior on $\rho$ is a uniform
in the interval $(-1, 1)$ which has a constant density in that interval.
The prior on $\gamma_1$, $\pi(\gamma_1)$, is $\pi(\rho) |\frac{\partial \rho}{\partial \gamma_1}|$, with

$$
\frac{\partial \rho}{\partial \gamma_1} = \frac{\partial (f(\gamma_1))}{\partial \gamma_1} = 2\frac{\exp(\gamma_1)}{(1 + \exp(\gamma_1))^2}
$$

Hence, the prior on $\gamma_1$ is

$$
\pi(\gamma_1) = \frac{1}{2} 2\frac{\exp(\gamma_1)}{(1 + \exp(\gamma_1))^2} =
  \frac{\exp(\gamma_1)}{(1 + \exp(\gamma_1))^2}.
$$

The prior on $\gamma_2$ is derived in a similar way, and it is

$$
\pi(\gamma_2) = \frac{\exp(\gamma_2)}{(1 + \exp(\gamma_2))^2}.
$$

\subsection{Impacts}

Computation of the impacts is done by exploiting that models are fit
conditional on $\rho$. For example, to compute the average total impact, the
posterior marginal of $\beta_r / (1 - \rho)$ is computed given $\rho$.  This is
easy as $\pi(\beta_r \mid \mathbf{y}, \rho)$ is estimated by INLA and the
posterior of $\beta_r / (1 - \rho)$ can be easily obtained by transforming the
posterior marginal of $\beta_r$.  Then, all the conditional marginals of the
average total  impacts are combined using BMA and associated weights $w_i$.
Posterior marginals for average direct and indirect  impacts can be computed in
a similar way.

\subsection{Final remarks}

It is worth stressing that model fit is done in parallel, which means that BMA
with INLA scale wells when the number of grid points or hyperparameters increases.
In our case, we have used function \texttt{mclapply} to fit all the models
required by BMA with INLA.

\texttt{R} code to run all the models shown in this paper with the different
methods is available
from GitHub at \url{https://github.com/becarioprecario/SAC_INLABMA}.

%%%%%%%%%%%%%%%%%%%%%%%%%%%%%%%%%%%%%%%%%%
\authorcontributions{This is a collaborative project. All authors contributed to the paper equally.}

%%%%%%%%%%%%%%%%%%%%%%%%%%%%%%%%%%%%%%%%%%
\funding{Virgilio G\'omez-Rubio was funded by 
Consejer\'ia de Educaci\'on, Cultura y
Deportes (JCCM, Spain) and FEDER, grant number SBPLY/17/180501/000491,
and by Ministerio de Econom\'ia y Competitividad (Spain), 
grant number MTM2016-77501-P.}

%Please add: ``This research received no external funding'' or ``This research was funded by NAME OF FUNDER grant number XXX.'' and  and ``The APC was funded by XXX''. Check carefully that the details given are accurate and use the standard spelling of funding agency names at \url{https://search.crossref.org/funding}, any errors may affect your future funding.}

%%%%%%%%%%%%%%%%%%%%%%%%%%%%%%%%%%%%%%%%%%
\acknowledgments{In this section you can acknowledge any support given which is not covered by the author contribution or funding sections. This may include administrative and technical support, or donations in kind (e.g., materials used for experiments).}

%%%%%%%%%%%%%%%%%%%%%%%%%%%%%%%%%%%%%%%%%%
\conflictsofinterest{The authors declare no conflict of interest.}

%%%%%%%%%%%%%%%%%%%%%%%%%%%%%%%%%%%%%%%%%%
%% optional
\appendixtitles{no} %Leave argument "no" if all appendix headings stay EMPTY (then no dot is printed after "Appendix A"). If the appendix sections contain a heading then change the argument to "yes".
%\appendix
%\section{}
%\unskip
%\subsection{}
%The appendix is an optional section that can contain details and data supplemental to the main text. For example, explanations of experimental details that would disrupt the flow of the main text, but nonetheless remain crucial to understanding and reproducing the research shown; figures of replicates for experiments of which representative data is shown in the main text can be added here if brief, or as Supplementary data. Mathematical proofs of results not central to the paper can be added as an appendix.
%
%\section{}
%All appendix sections must be cited in the main text. In the appendixes, Figures, Tables, etc. should be labeled starting with `A', e.g., Figure A1, Figure A2, etc. 

%%%%%%%%%%%%%%%%%%%%%%%%%%%%%%%%%%%%%%%%%%
\reftitle{References}

\externalbibliography{yes}
\bibliography{biblio}

\end{document}